\documentclass[12pt,preprint]{aastex}
\newcommand{\beq}{\begin{equation}}
\newcommand{\eeq}{\end{equation}}
\newcommand{\bdm}{\begin{displaymath}}
\newcommand{\edm}{\end{displaymath}}

\slugcomment{Submitted to the Astrophysical Journal (Letters)}

\received{2003 April 21}
\begin{document}

\title{A Compact X-ray Source and Possible X-ray Jets within the
  Planetary Nebula Menzel 3} 

\author{Joel H. Kastner\altaffilmark{1}, Bruce Balick\altaffilmark{2}, Eric G. Blackman\altaffilmark{3}, Adam Frank\altaffilmark{3},
Noam Soker\altaffilmark{4}, Saeqa
D. Vrt\'ilek\altaffilmark{5}, and Jingqiang Li\altaffilmark{1}}

\altaffiltext{1}{Chester F. Carlson Center for
Imaging Science, Rochester Institute of Technology,
Rochester, NY 14623; jhk@cis.rit.edu}
\altaffiltext{2}{Department of Astronomy, University of
Washington, Box 351580, Seattle, WA  98195, USA;
balick@astro.washington.edu}  
\altaffiltext{3}{Department of Physics and Astronomy,
University of Rochester, Rochester, NY 14627, USA}
\altaffiltext{4}{Department of Physics, Oranim, Tivon 36006, Israel; 
soker@physics.technion.ac.il}
\altaffiltext{5}{Harvard-Smithsonian Center for
Astrophysics, Cambridge, MA 02138, USA; saku@cfa.harvard.edu}

\begin{abstract}

We report the discovery, by the Chandra X-ray Observatory,
of X-ray emission from the bipolar planetary nebula Menzel
3. In Chandra CCD imaging, Mz 3 displays hot ($\sim3-6\times10^6$
K) gas within its twin, coaxial bubbles of optical nebulosity,
as well as a compact X-ray source at the position of its
central star(s). The brightest diffuse X-ray emission lies
along the polar axis of the optical nebula, suggesting a
jet-like configuration. The observed combination
of an X-ray-emitting point source and possible X-ray jet(s)
is consistent with models in which accretion disks and, potentially,
magnetic fields shape bipolar planetary nebulae via the
generation of fast, collimated outflows.

\end{abstract}

%\keywords{stars: AGB and post-AGB --- stars: mass loss ---
%stars: winds, outflows --- ISM: jets and outflows ---
%planetary nebulae: individual (Menzel 3)}

\section{Introduction}

Planetary nebulae (PNs) are the descendents of intermediate-mass
stars (initial masses $\sim1-8 M_\odot$).
The central star of a PN eventually will 
evolve into a white dwarf, with a mass of between $\sim0.55
M_\odot$ and $\sim1.0 M_\odot$. Much of the star's initial mass
is expelled during the $\sim10^5$ years that the star spends on
the upper asymptotic giant branch (AGB). This AGB
star wind usually consists of a more or less spherically
symmetric outflow at rates of up to $10^{-4}$ $M_\odot$
yr$^{-1}$ and speeds of $\sim10-20$ km s$^{-1}$.  
As the star leaves the AGB, however, its
surface escape velocity and, hence, wind speed, drastically
increases to $\sim1000$ km s$^{-1}$ and its mass loss rate drops to
$10^{-8}$ $M_\odot$ yr$^{-1}$. Eventually, the UV flux from this emerging
white dwarf ionizes the AGB ejecta, producing a
PN detectable in atomic recombination line emission.

Optical images of PNs obtained by the Hubble Space Telescope
(HST) very clearly show that, during late AGB and/or early
PN stages, the geometry of the outflow often takes on axial
(bipolar) or even more spectacular, high-order symmetries
(Sahai \& Trauger 1998; Kastner, Soker \& Rappaport 2000a;
Balick \& Frank 2002). Various models have been proposed to
explain this transition; some or all of these may apply at
different times and to various central star configurations.
For example, during late AGB phases when the mass loss rate
is highest, the accretion of material by a nearby companion
may cause a collimating disk to form (Mastrodemos \& Morris
1999; Soker \& Rappaport 2000; and references therein). The
presence of such a companion and disk potentially provides
an additional source of wind momentum and/or ionizing UV
flux, thereby significantly altering the ``standard''
picture of late-AGB, early-PN evolution described
above. Another proposal is that magnetic fields emerge from
the stellar core as the AGB star's outer layers are stripped
away (Blackman et al. 2001b).  The resulting $\sim1$ G
fields can either help to collimate the radiation driven
wind, or mediate a stronger, more anisotropic, magnetically
driven wind. This stellar wind can interact with, and be
collimated by, a surrounding magnetically mediated disk
wind.  Disk winds and stellar winds may therefore operate in
tandem to produce the observed rates and morphologies of
mass loss and very high momenta of outflowing material prior
to the onset of wind ionization by the emerging white dwarf
(Blackman et al.\ 2001a, Frank et al.\ 2003).

The tenuous but very fast stellar
wind from the emerging central white dwarf (or a companion,
if the mass-losing central star is still on the AGB)
will shock the relatively slow-moving AGB ejecta, creating an 
overpressured (and therefore rapidly expanding) ``hot
bubble'' interior to the optical nebula. The shocked gas 
generated by these PN wind interactions should be detectable
in the form of X-ray emission that fills an optically dark
cavity within swept-up stellar ejecta.  
Simple energy conservation arguments predict that such a wind-heated
bubble can reach temperatures of $10^7-10^8$ K, although it is
likely that the bulk of the shocked gas will appear cooler
($\sim10^6$ K), due to various processes (such as early wind
onset, adiabatic expansion, and conduction of heat from the wind-shocked gas
to cooler, denser nebular gas) that can moderate the gas
temperature (Soker \& Kastner 2003a, and references
therein). In addition, the new classes of PN shaping models that rely on
magnetohydrodynamics and/or the presence of accreting,
magnetically active companion stars predict magnetic
fields at the base of the flow that are strong enough to
produce detectable, compact X-ray emission (Blackman et al.\
2001a; Soker \& Kastner 2002). Hence, there
is reason to expect both point-like and diffuse X-ray
emission from some PNs.  

Correspondingly, recent observations by the Chandra X-ray
Observatory (CXO) and XMM-Newton have revealed examples of
diffuse, X-ray-emitting gas within the well-studied PNs BD
$+30^\circ$3639 (Kastner et al.\ 2000b), NGC 7027 (Kastner,
Vrt\'ilek \& Soker 2001), and NGC 7009 (Guerrero et al.\ 2002),
point-like emission within NGC 7293 (Guerrero et al.\ 2001),
and both diffuse and point-like emission within NGC 6543
(Chu et al.\ 2001; Guerrero et al.\ 2001).  In optical and
infrared images, all four PNe thus far detected in diffuse
X-ray emission share a general elliptical morphology,
punctuated by knots or loops indicative of recently
established, high-velocity, collimated flows.
In this {\it Letter} we extend the high-resolution X-ray
observations to the important class of extreme bipolar PNe
(referred to as ``butterfly'' PNs; Balick \& Frank 2002), via
CXO observations of Menzel 3 
(Mz 3 = Hen 2-154 = PK 331-01 1 = PN G331.7-01.0). 

Mz 3 has long been considered a prototypical example of an
extreme bipolar PN (L\'opez \& Meaburn 1983). Like many
objects in this class, 
however, the core of Mz 3 is evidently a symbiotic Mira system,
consisting of an AGB primary with dense, slow winds and a 
companion white dwarf that provides the UV flux which ionizes the
nebula. The symbiotic Mira nature of Mz 3 is revealed by deep
optical spectra (Zhang \& Liu 2002) and by
infrared colors (Schmeja \& Kimeswenger 2001). If Mz 3 is
indeed a symbiotic system, then it may not be a PN, but rather
a PN in the making; nevertheless,
we refer to it as a PN in this paper.

In high-resolution, visual-wavelength images obtained by the
HST, Mz 3 presents a highly complex
morphology (Fig.\ 1). Its bright, bipolar lobes appear
as enclosed bubbles embedded within a larger cylindrical structure;
this structure is in turn surrounded by an extended set of
``streamers'' that seem to point directly away from the
central star. The
distance to Mz 3 is poorly known (Zhang \& Liu
2002) but, adopting an estimate of 2.7 kpc (Kingsburgh \&
English 1992), each enclosed bipolar bubble is $\sim$40,000
AU long. 

High-resolution optical spectroscopy reveals that gas in Mz 3
is outflowing along its polar axis at
velocities of between 300 km s$^{-1}$ and 500 km s$^{-1}$ (Redman et
al.\ 2000; Balick, unpublished; Guerrero et al.\ 2003). Such
velocities are only observed, for PNe, in the case of the
most extreme bipolar objects (Bujarrabal et al.\ 2001). The
potential for X-ray-generating shocks due to the collimated,
fast winds in Mz 3, and the evidence that it harbors a
central binary system, makes this nebula a
particularly interesting object to probe for X-ray emission at
high resolution.

\section{Observations}

We observed Mz 3 with Chandra for 40.83 ks on 2002 October
23. The detector was the back-illuminated CCD S3 of the
Advanced CCD Imaging Spectrometer (ACIS).  ACIS has a pixel
size of 0.49$''$ --- very similar to the width of the point
spread function (PSF) of the Chandra mirrors --- and the
Chandra/ACIS combination is sensitive over the energy range
0.3-10 keV. The data, consisting of individual CCD X-ray and
particle events, were subject to standard processing by
Chandra X-ray Center pipeline software
(CIAO\footnote{http://cxc.harvard.edu/ciao/}, version 
2.2), which
determines the distribution of photon-generated charge
within a 3$\times$3 CCD pixel island centered on the event
position, flags events likely due to particles, and computes
the celestial positions and nominal energies of incident
X-rays. To optimize image resolution, the event charge
distibutions were used, along with the telescope pointing
history and nominal photon energies, to calculate a subpixel
position for each X-ray (Li et al.\ 2003).

\section{Results}

The Chandra/ACIS-S3 image of Mz 3 (Fig.\ 1)
represents the first detection of X-rays from 
this object.  Both diffuse and point-like X-ray emission is
detected. The central X-ray
source lies at the position of the central star seen in HST
images, based on the spatial coincidence of several
background HST sources with X-ray sources in the Chandra/ACIS-S3 field.

We used CIAO tools (version 2.3) to extract X-ray source
counts and spectra within $7.5''\times19''$ elliptical and
$3.5''$ diameter regions
encompassing the diffuse and point-like emission,
respectively. We find a total, background-subtracted count rate of
1.9$\pm$0.2 ks$^{-1}$ ($0.2-2.5$ keV) from the combined diffuse and
point-like components. We estimate that $\sim60$ total counts arise from the
diffuse emission, and $\sim17$ from the central source.

Nearly all of the detected emission from Mz 3 is confined to
energies between 0.6 and 2.0 keV (Fig.\ 2). Its X-ray
spectral distribution is similar to that of NGC 7027,
indicative of an X-ray emission temperature in the range
$3-8\times10^6$ K (Kastner et al.\ 2001; Maness et al.\
2003). To confirm this, we used CIAO spectral fitting tools
to perform fits of a variable-abundance thermal emission
(VMEKAL) model with intervening absorption. Results from
optical spectroscopy (Zhang \& Liu 2002, their Table 11)
were used to constrain the model abundances. Adopting these
abundances, we find an X-ray emission temperature of $T_x =
6\times10^6$ K and absorbing column of $N_H =
5\times10^{21}$ cm$^{-2}$. Because the observed spectral
energy distribution is sharply peaked, the formal
uncertainties in these parameters are $\sim\pm15$\%, despite
the poor photon counting statistics. Given the likelihood
that the abundances of the X-ray-emitting gas depart
significantly from the nebular abundances (e.g., Chu et al.\
2001; Maness et al.\ 2003), however, $T_x$ and $N_H$ are
actually not so well constrained, and the model fitting
procedure cannot be used to further refine abundances for
the X-ray emitting region (in contrast to results in, e.g.,
Maness et al.). In particular, although the best-fit model
abundances of Ne and Fe --- 0.4 and 0.2 solar, respectively
--- appear consistent with the nebular values (0.4 and 0.3 solar,
respectively; Zhang \& Liu), each is uncertain by a
factor $\sim3$, reflecting the uncertainty in assigning the
emission peak near 0.9 keV to highly ionized Ne, highly
ionized Fe, or both. If Fe is severely depleted (as appears to
be the case for BD $+30^\circ$ 3639; Maness et al.\
2003), we find that values of $T_x \sim 3\times10^6$ K and
$N_H \sim 8\times10^{21}$ cm$^{-2}$ provide the best fit.

Adopting values of $T_x = 6\times10^6$ K and $N_H = 5\times10^{21}$
cm$^{-2}$, the absorbed flux derived from the model fitting procedure
is $7\times10^{-15}$ erg cm$^{-2}$ s$^{-1}$. The implied total
(combined diffuse and point-like) intrinsic source X-ray luminosity is
$\sim3\times10^{31}$ $(\frac{D}{2.7 {\rm kpc}})^2$ erg
s$^{-1}$, where $\sim6\times10^{30}$ $(\frac{D}{2.7 {\rm kpc}})^2$ erg
s$^{-1}$ can be ascribed to the point source under the
assumption that its intrinsic spectrum is not much harder
than that of the diffuse component (see below).

As in previous X-ray detections of PNs, the modeling
results for intervening X-ray absorbing column 
are consistent with measurements of visual extinction. Zhang
\& Liu (2002) deduce $E(B-V) = 1.38$ and $E(B-V) = 1.59$ toward
the nebula and the central star, respectively, suggesting
$A_V \approx 4.5$. Hence, the ratio $N_H/A_V$ in Mz 3 ($\sim 10^{21}$
cm$^{-2}$ mag$^{-1}$) is similar to that in BD +30$^\circ$3639 and
in NGC 7027 (Kastner et al.\ 2000b, 2001, 2002; Maness et
al.\ 2003). There is probably a significant absorption
component that is due to the interstellar medium along the
line of sight to Mz 3, given its large estimated
distance and low galactic latitude ($b = -1.01^\circ$). 
The nebula is a strong far-infrared
source (IRAS 25 $\mu$m flux of 343 Jy) and displays a local
component of diffuse optical absorption bands (Zhang
\& Liu), however, indicating that some or
most of the soft X-ray absorption 
is due to dense gas and dust within the nebula itself.

\section{Discussion}

\subsection{An X-ray jet and obscured central X-ray source in Mz 3}

The CXO imaging reveals a strong correspondence between the diffuse
X-ray emitting region of Mz 3 and its visual-wavelength emission
(Fig.\ 3).  Specifically, the diffuse X-ray emission is confined to
the interiors of the bubbles of ionized gas seen in HST imaging. This
correspondence resembles those of earlier CXO and XMM-Newton
detections of diffuse X-ray emission from planetaries
(Kastner et al.\ 2000b, 2001, 2002; Chu et al.\ 2001; Guerrero et al.\
2002). With the possible exception of NGC 7027 (Kastner et
al.\ 2001, 2002), the X-ray emitting gas is precisely outlined by 
bright elliptical shells within these nebulae. 

In contrast, the brightest X-ray emission from Mz 3 lies
close to the symmetry axis of the object (Fig.\ 3). This
morphology strongly suggests that most of the diffuse X-ray
emission is associated with collimated outflows, perhaps in
the form of jets. Jet-like structure is suggested by the
diffuse emission south of the nucleus. In particular, there is
a ``knot'' of emission located along the polar axis,
$\sim11,000$ AU ($4''$) south of the source. Based on the
density of field X-ray sources within $\sim1'$ of Mz 3, we
estimate that there is only a $\sim3$\% probability that
this feature is a foreground or background X-ray
source. A plausible interpretation of the diffuse X-ray
morphology within the south lobe of Mz 3, therefore, is that
this lobe harbors a ``blobby'' X-ray jet that extends at
least 18,000 AU ($7''$) from the central source, perhaps
terminating at the leading edge of the southern lobe.

To the north of the central star the X-ray and optical
surface brightness distributions are somewhat more
amorphous, with the X-ray emission brightest near the tip
of, and apparently outlining, the protrusion in the north
lobe. No feature corresponding to the southern X-ray ``knot''
is seen immediately to the north of the central
star. Instead, there appears to be a gap within the diffuse
X-ray nebula, spatially coincident with a similar feature in
optical recombination line emission, just north of the
central source.  As high-resolution optical spectroscopic
observations indicate that the north lobe is slightly tipped
away from the observer (e.g., Redman et al.\ 2000), it
is possible that the lack of emission just to
the north of the central star is due to the
orientation of the nebula, which is such that the inner
northern lobe is partially obscured by material located
along the equatorial plane. Alternatively, 
the X-ray emission along the polar axis of Mz 3 may be
intrinsically asymmetric, as is the
case in another symbiotic Mira system, R Aqr (Kellogg
et al.\ 2001; see \S 4).

All of the photons detected from the central
source have energies $\ge 1$ keV, suggesting either that
this source is intrinsically harder than the shocked nebular
gas or, perhaps, that the central source is especially deeply
embedded in X-ray absorbing gas. The latter interpretation
is consistent with the relative degree of visual extinction,
which is larger toward the central star than toward the
nebula (Zhang \& Liu 2002).

\subsection{The X-ray evidence for magnetized accretion
disks in symbiotic Miras}  

The combination of an X-ray-bright core and X-ray emission along
the polar axis of Mz 3 resembles that observed in R Aqr (Kellogg et
al.\ 2001), which is also a symbiotic Mira system. Adopting
respective distances of 2.7 kpc and 200 pc, however, the
X-ray emission from Mz 3 appears much more luminous and
extensive than that of R Aqr, which has a total X-ray
luminosity $\sim10^{30}$ erg s$^{-1}$ and whose
discontinous, southwest jet (the longer of its two,
asymmetric X-ray jets) extends $\sim6000$ AU from the
central source. If, instead, Mz 3 is as close as $\sim1$ kpc
--- the smallest of many, disparate distance estimates for
this PN (see Zhang \& Liu 2002 and references therein) ---
the X-ray properties of this symbiotic Mira system and R Aqr
would be remarkably similar.

Alternative explanations for the unresolved, core X-ray
emission within Mz 3 and R Aqr (see also Guerrero et al.\
2001 and Soker \& Kastner 2002) are (1) magnetic
star-disk interactions (or, perhaps, 
accretion shocks) associated with accretion by a companion
to the central AGB star; (2)
star-disk interactions or shocks due to accretion
by the AGB star itself 
(if the secondary is less than a solar mass, and approaches
close enough to the AGB star to be tidally disrupted;
Reyes-Ruiz \& L\'opez 1999); (3) stellar magnetic activity
originating with the AGB star (e.g.,
Blackman et al.\ 2001a; Garc\'ia-Segura, L\'opez, \& Franco
2003; Soker \& Kastner 2003b); (4)
X-ray-generating wind shocks analogous to those around O/B
stars (e.g., Cassinelli et al.\ 1994); or, perhaps most
speculatively, (5) hard X-ray emission from 
the companion white dwarf itself (e.g., O'Dwyer et al.\ 2003).

Given that both Mz 3 and R Aqr evidently harbor symbiotic
binary central stars, it seems most likely that the compact
X-ray source arises, in each case, from an accretion disk
around a hot secondary (alternative 1, above). By analogy
with, e.g., models for disk-driven winds from young stellar
objects (YSOs), such a configuration also would explain the
presence of high-velocity, collimated winds that generate
X-rays as they collide with ambient nebular
gas (e.g., Pravdo et al.\ 2001). 

It has been proposed that magnetic fields play a
fundamental, dynamical role in disk-jet systems as diverse
as YSOs (e.g., Pudritz \& Konigl 2000), active galactic
nuclei (AGNs; e.g., Ferrari 1998 and references therein),
and, indeed, R Aqr (Hollis \& Koupelis 2000). The
combination of collimated, ``knotty,'' X-ray emitting outflows and
compact, central X-ray source observed in both Mz 3 and R
Aqr is, morphologically, similar to that characteristic of
AGNs (e.g., 3C 273, Marshall et al.\ 2001; Cen A, Hardcastle et
al.\ 2003).  In the case of the
symbiotic Mira systems --- as in AGNs --- a natural relation
between the X-ray-emitting core and jet(s) would be that the
central engine has a mixture of closed small scale and open
large scale fields, dynamically reforming. The jets flow
along open field lines, while the point-like X-ray emission
is generated via the dissipation of the closed structures as
they open, much like solar flares. This suggests that the
core X-ray emission in Mz 3 and R Aqr may be variable;
sensitive X-ray monitoring of these nebulae (and of NGC 6543, the other
known example of both diffuse and point-like X-ray emission
in a PN) is required to establish whether or not this is the case.

We caution that it remains to establish theoretically, via detailed
numerical modeling, whether magnetic fields necessarily play
a central role in either the launching or the collimation of
jet-like outflows from evolved star systems. Nonetheless,
the presence of compact X-ray sources and the appearance of
X-ray emission that is confined to the outflow axes
of the symbiotic Mira systems 
Mz 3 and R Aqr supports models that invoke collimating
circumstellar disks driving high-velocity jets to explain
the profound bipolar structure of these and other, similar
nebulae around evolved stars.

\acknowledgements{Support for this research was
provided by NASA/CXO grant GO2--3009X to RIT. Geoffrey
Franz assisted in preparation of data for Figs.\ 1 and 3.
N.S. acknowledges support from the US-Israel
Binational Science Foundation. S.D.V acknowledges support
from NASA grant NAG5--6711. }

\clearpage
\begin{figure}[htb]
\includegraphics[scale=1.,angle=0]{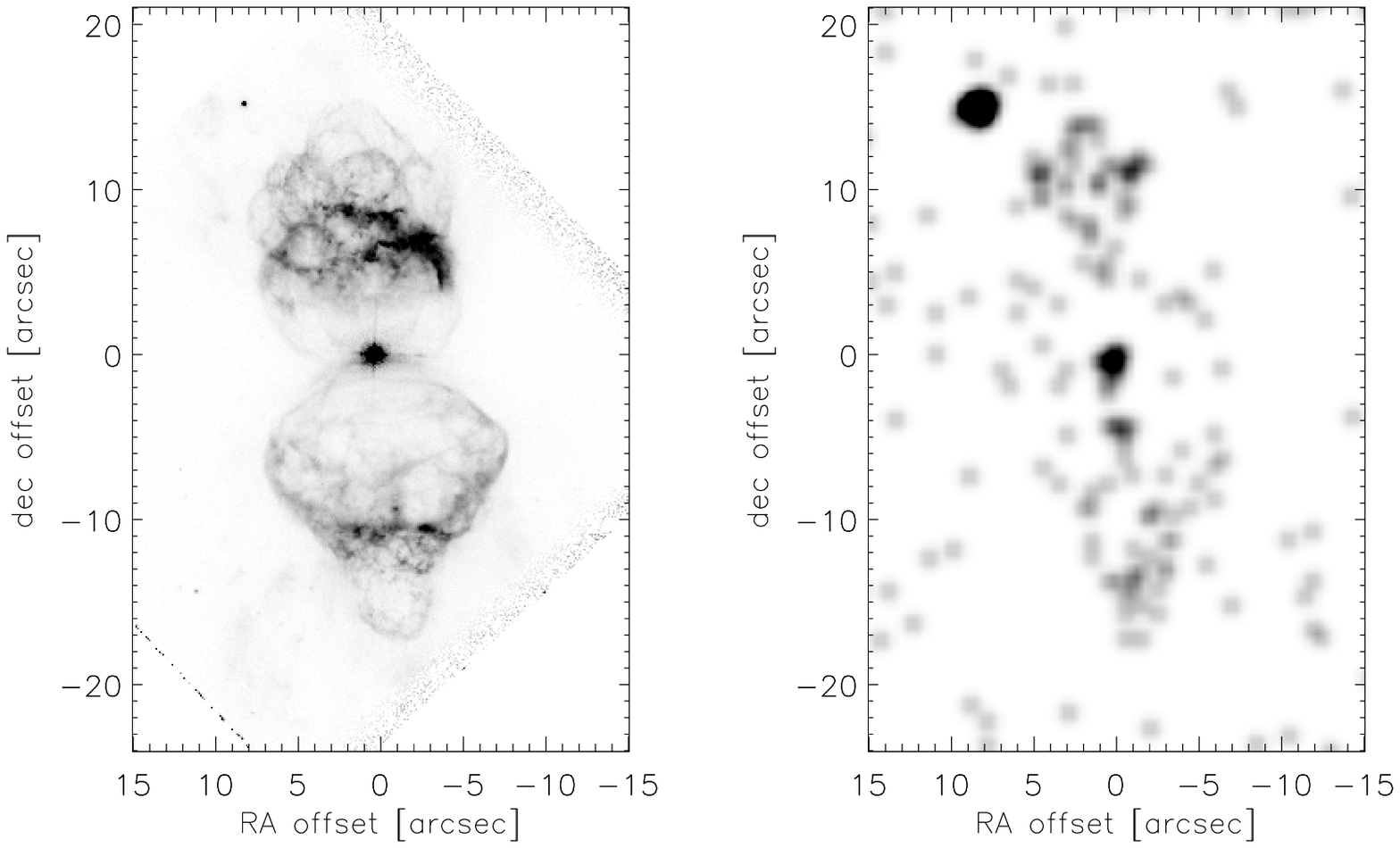}
\caption
{Comparison of HST visual-wavelength (H$\alpha$; left) and
Chandra/ACIS-S3 X-ray (right) images of Menzel 3. The X-ray
image is constructed for the energy range 0.7 to 2.0 keV.}
\end{figure}

\begin{figure}[htb]
\includegraphics[scale=1.,angle=0]{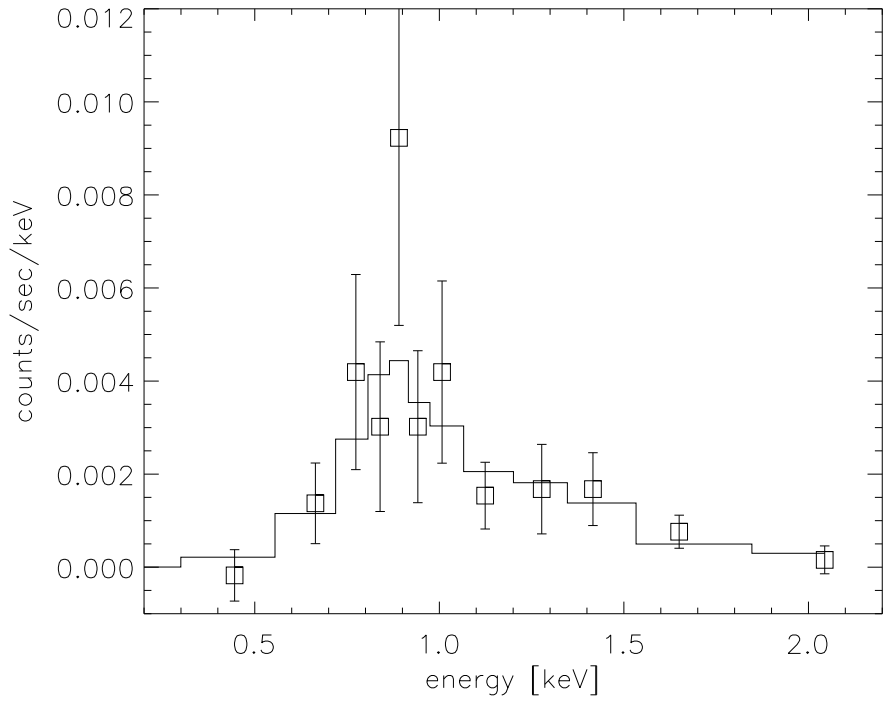}
\caption
{ACIS-S3 spectrum of Menzel 3 (squares), with best-fit
absorbed VMEKAL emission model (histogram) overlaid.}
\end{figure}

\begin{figure}[htb]
\includegraphics[scale=1.,angle=0]{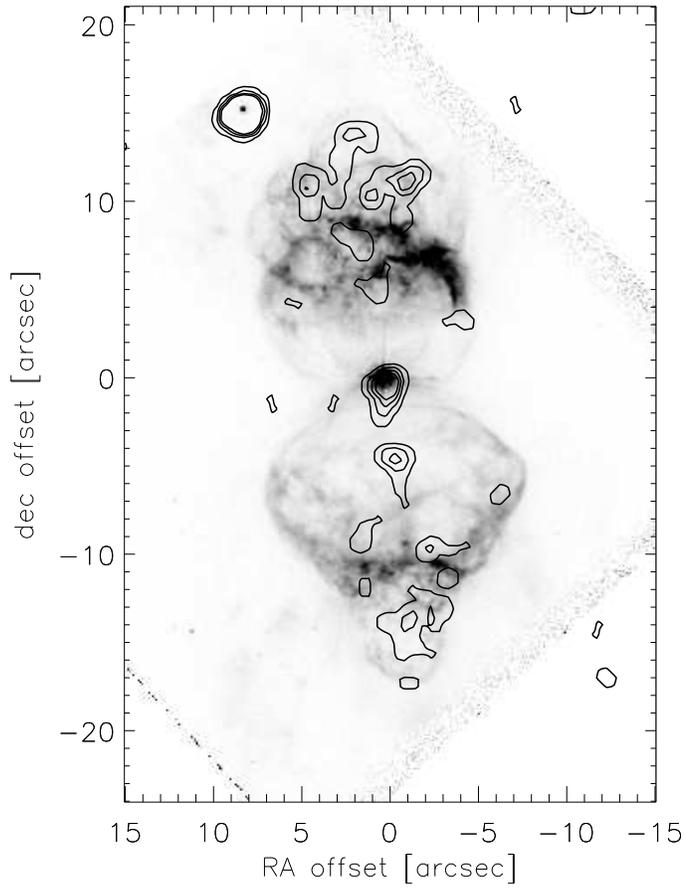}
\caption
{HST H$\alpha$ image of Menzel 3, with contour plot of
Chandra/ACIS-S3 image overlaid. Contours are at levels of
1.0, 2.0, 3.0, and 4.0 counts pixel$^{-1}$.}
\end{figure}

\end{document}